\documentclass[%
 pre,showpacs,showkeys,10pt,twocolumn, 
nofootinbib,
 amsmath,amssymb,
 aps,
]{revtex4-2}

\usepackage{graphicx}
\usepackage{dcolumn}
\usepackage{bm}

\begin{document}
	
\title{Combinatorial Calculations of the Most Probable Short‑term Return of Single Company Stocks on the Stock Market}

\author{Michael Romanovsky}
\email[]{slon@kapella.gpi.ru}

\affiliation{PE Science and Innovationshaya Ordynka Street, Moscow, Russian Federation \\ ANO National Center for Physics and Mathematics, ul. Parkovaya 1, str. 3, 607182 Sarov, Nizhny Novgorod region, Russian Federation \\ Pirogov Russian National Research Medical University, ul. Ostrovityanova 1, 117997 Moscow, Russian Federation }

\date{\today}

\begin{abstract}
The combinatorial method is used to calculate the quantity of distribution variants for the number of stockholders of one individual company as well as the quantity of stocks in the stockholder portfolio. These quantities may change as a result of the stocks buy and sale that is taken into account also. The total number of stocks — the units at the lower level of the problem — is obviously preserved, while the number of portfolios — the structural units at the upper level — may change.

The most probable distributions of the number of portfolios and stocks in each of them are determined depending on the expense to add a stock to the portfolio (the purchase price). The application of the developed method to determine the most probable short‑term return on stocks in a combinatorial sense is demonstrated.

\end{abstract}



\maketitle

\section{Introduction}

The number of possible distribution variants of $N$ stocks of a single selected company that are traded on the stock market among the portfolios of their holders at a certain point in time \footnote{We assume that this period does not coincide with the dividend period, and we do not distinguish between preferred and ordinary shares.}  is characterized at its maximum, in the absence of any obstacles to exchange, by the number $N!$. This number arises because the replacing of one stock with another in the initial “chaos” results in a new distribution. At the same time, there are obstacles prevent transitions of stocks between the holders. They are the non-zero value of the stocks, as well as the physical and legal accessibility of the portfolios of their holders. These circumstances reduce the number of possible distributions. “Inside” each {\it i}-th portfolio, there are $(p_i)!$ possible ways to distribute the stocks; this fact reduces the number of possible distribution variants by the value of ${\prod}_{i=1}^m(p_i)!$, where $m$ is the total number of different stockholders (the number of portfolios), and $p_i$ is the number of stocks in the {\it i}-th portfolio. In addition, the number $N!$ is reduced by the number of variants that led to the distribution of $N$  stocks across $m$ portfolios — by the number of placements with repetitions, $m^N$.

Thus, the total number of possible distributions of $N$ stocks across $m$ portfolios, taking into account the number of their distributions inside the portfolios and the placement when forming portfolios, is

\begin{equation}
	W = {N! \over{m^N {{\prod}_{i=1}^m}{(p_i !)}}}
	\label{eq_1}
\end{equation}
It is clear that

\begin{equation}
	N = {{\sum}_{i=1}^m}{p_i}
	\label{eq_2}
\end{equation}

The concentration of all stocks in the single owner hands (this usually does not happen for stocks traded on the stock exchange) corresponds to $m = 1$, while the complete dispersion of portfolios containing a single stock corresponds to $m = N$ (this also does not happen). Typically, $m$ varies within a wide range, but it still remains large. The problem is to determine the dependence of the price on $m$ and/or $p_i$. Combinatorial analysis allows us to determine the most probable value of $W$ depending on $m$ and $p_i$. This combinatorial problem, which is supplementary to those considered in ~\cite{isihara1971statistical,romanovsky2021statistical, romanovsky2022chaos, romanovsky2026UFN}, belongs to the class of combinatorial problems involving the calculation of the number of particle combinations across objects under various constraints. The total number of possible combinations of $N$ such particles $W$ is determined by the formula

\begin{equation}
	W = {N! \over{F([m],[p_i])}}
	\label{eq_3}
\end{equation}
where the function $F([m],[p_i])$  will be determined through various combinatorial operations with sets of particles $[p_i]$ in a set of $[m]$ higher-level structural units. Various cases of the implementation of the function $F([m],[p_i])$, including those more complex than (\ref{eq_1}), are considered in ~\cite{romanovsky2026UFN}. 

\section{The combinatorial problem of determination the most probable distributions of the number of holders’ portfolios and the stock parts in them.}

To determine the most probable distribution, one should first, as usual ~\cite{isihara1971statistical}, switch in (\ref{eq_1}) from factorials to their expression via Stirling’s formula (it is enough to use it in the first approximation), and then calculate the variations of the expression obtained from (\ref{eq_1}) with respect to $m$ and all $p_i$, taking into account the Lagrange terms. We have only one Lagrange multiplier term, since only the total number of stocks — structureless particles $N$ (\ref{eq_2}) — is conserved, and it is equal to ${\alpha} N$, where ${\alpha}$ is the Lagrange multiplier. Usually, the quantity $ln W_1={\ln}W+{\alpha}N$ is varied directly:   

\begin{equation}
ln W_1 = N(ln N -1)- N ln m - {{\sum}_{i=1}^m}{p_i}({\ln}p_i - 1) + {\alpha} N
	\label{eq_4}
\end{equation}
By varying (\ref{eq_4}) over $p_i$ using (\ref{eq_2}), we obtain the corresponding derivative, which we set equal to zero:

\begin{equation}
{{{\partial}{ln W_1} \over {{\partial} p_i}}} = ln N - ln m - ln p_i + {\alpha} N = 0
	\nonumber
\end{equation}
or

\begin{equation}
m p_i = N e^{\alpha}
	\label{eq_5}
\end{equation}

where we used

\begin{equation}
{{{\partial}{N} \over {{\partial} p_i}}} = {{{\partial}\over {{\partial} p_i}} {\sum}_{i=1}^m}{p_i} = 1 
	\nonumber
\end{equation}
Thus, we have obtained the first equation of the system.

To obtain the second equation of the system from the variation over $m$, it is necessary to determine the derivative of the sum of some quantities by its upper limit 

\begin{equation}
{{{\partial} \over {{\partial} m}} {\sum}_{i=1}^m}{f_i}
	\nonumber
\end{equation}

Recalling the well-known expression for the derivative of a definite integral over the upper limit:

\begin{equation}
{{{\it d} \over {{\it d} x}} {\int}_a^x}{f(y)d y } = f (x), 
	\nonumber
\end{equation}
let’s define the derivative of the sum of some quantities $f_i$ by its upper limit by analogy

\begin{equation}
{{{\partial} \over {{\partial} m}} {\sum}_{i=1}^m}{f_i} = f_m
	\label{eq_6}
\end{equation}
where $f_m$ is the last value of the summed values. However, in the future, bearing in mind that all $p_i$, generally speaking, are basically the same functions, and using the independence of the sum from the place of the summand in it, we will assume

\begin{equation}
{{{\partial} \over {{\partial} m}} {\sum}_{i=1}^m}{p_i} =p_i
	\nonumber
\end{equation}
Then

\begin{equation}
{{{\partial}{ln W_1} \over {{\partial} m}}} = p_i (ln N - ln m - ln p_i + {\alpha} N) - {N \over m} + N {{{\partial}{ln N} \over {{\partial} m}}}
	\label{eq_7}
\end{equation}
Let us calculate the derivative ${{{\partial}{ln N} \over {{\partial} m}}}$ using rule (\ref{eq_6}) and the definition of the derivative

\begin{equation}
{{{\partial}{ln N} \over {{\partial} m}}}= {{{\partial} \over {{\partial} m}} ln ({{\sum}_{i=1}^m}{p_i}) = {{ln ({{\sum}_{i=1}^m}{p_i}) - ln ({{\sum}_{i=1}^{m-1}}{p_i})}  \over {m - (m-1)}}}
	\nonumber
\end{equation}
We get the second equation in the form (again, we replace $p_m$ with $p_i$)

\begin{equation}
{{{\partial}{ln N} \over {{\partial} m}}}= - ln (1- {p_i \over N})
	\nonumber
\end{equation}

Substituting this expression into (\ref{eq_7}) and remembering that the expression in parentheses in (\ref{eq_7}) equals zero due to (\ref{eq_5}), we finally obtain the second equation of the system:

\begin{equation}
ln (1- {p_i \over N}) + {1 \over m} = 0
	\label{eq_8}
\end{equation}
The system of equations (\ref{eq_5}) and (\ref{eq_8}), taking into account that has been said about $p_m$, solves the problem. Let's write it down in its final form
\begin{equation}
m p_i = N e^{\alpha},
\nonumber \\ 
\end{equation}
\begin{equation}
{ln (1- {{p_i}  \over N})} + {1 \over m} = 0
	\label{eq_9}
\end{equation}
The variables $m$ and $p_i$ can be elementary decoupled, then

\begin{eqnarray}
{N \over p_i} ln (1- {p_i \over N}) + e^{- \alpha} = 0
 \nonumber \\ 
m (e^{- {1 \over m}} - 1)+ e^{\alpha} = 0
	\label{eq_10}
\end{eqnarray}

It is obvious that in this notation, the variables $m$ represent the number of stockholder portfolios, and ${p_i \over N}$ represents the quantity part of the company’s stocks in the $\it i-$th portfolio. Note that the first terms in both equations (\ref{eq_10}) are always negative. 

It also immediately follows from the second equation (\ref{eq_10}) that $\alpha$ is negative. If we interpret $\alpha$ as a quantity proportional to the work required to detach a stock from the portfolio, i.e., to sell it, while keeping the total number of portfolios $m$ constant (${\alpha}_0$ is proportional to the stock price before the sale, ${\alpha}_1$ - after the sale), and use the firts equation (\ref{eq_9})

\begin{equation}
m (p_i + 1) - m p_i = N (e^{{\alpha}_1} - e^{{\alpha}_0}) = m > 0
	\nonumber
\end{equation}
Then ${\alpha}_1 < {\alpha}_0 < 0$, i.e., as expected, and the stock price under the sale $\sim e^{{\alpha}_1} - e^{{\alpha}_0}$ should be positive.

In the above calculations, $\alpha$ is definitely a dimensionless quantity, proportional, as stated, to the work of attaching/detaching stocks to/from the portfolio, i.e. the price. This value contains a normalization value with the dimension of a certain price. In such combinatorial problems (see, for example, ~\cite{romanovsky2021statistical, romanovsky2022chaos}), this normalization is usually referred to as the effective temperature (of total stock market) $T_{eff}$. In our problem, it again has the price dimension. We will see later that some conclusions can be drawn from the equations obtained to solve our given problem even without determining $T_{eff}$ — by simply indicating what limitations are imposed on it, and these limitations will be also the limitations for the problem being addressed.

\section{Determination of the most probable stock return over a limited time interval}

By definition, the stock return over the time interval $\Delta t$ from time $t$ to time $t + \Delta  t$ is

\begin{equation}
Y = ln ({c_{t + \Delta t} \over c_t}) ,
	\label{eq_11}
\end{equation}
where $c_t$ is the stock price at time $t$. In the previous section, we determined that the most probable value $\alpha \sim c$, that is the stock price at a given specific moment, i.e., ${\alpha}_t = {c_t \over {T_{eff} (t)}}$, and therefore ${{\alpha}_{t + \Delta t}} = {c_{t + \Delta t} \over {T_{eff} (t + \Delta t)}}$. Then

\begin{equation}
Y = ln [{{\alpha}_{t + \Delta t} \over {\alpha}_t} {T_{eff}(t + \Delta t) \over T_{eff}(t)}] ,
	\label{eq_12}
\end{equation}

The stock return is a rapidly changed and very often time–varying quantity. According to all modern concepts ~\cite{romanovsky2011vidov, mantegna2000stanley}, it is delta-correlated over time. In turn, $T_{eff}$ is obviously related to global stock market processes, i.e. it can be expected to change much more slowly over time. In principle, it is generally possible to define $T_{eff}$ as some effective normalization of the stock price in the market, and put it constant. However, this approach completely ignores the market's influence on the stock price, which is incorrect. 

It is more correct to use the opportunity to define $T_{eff}$ really as some "effective temperature" of the stock market, which also has a price dimension. Temporary changes in this "effective temperature" are most obviously related to temporary fluctuations in the stock market volatility, which is determined for the stock price as

\begin{equation}
G_c = | ln ({c_{t + \Delta t} \over c_t}) |.
	\nonumber
\end{equation}

For the entire market, they usually use the volatility value of a specific stock index $S$ (for example, $SP500$) $G_S$:

\begin{equation}
G_S = | ln ({S_{t + \Delta t} \over S_t}) |.
	\nonumber
\end{equation}

The autocorrelations of both $G_c$ and $G_S$ decrease significantly over a very long period of time typically, that is measured in months ~\cite{romanovsky2020romanovsky, stanley1999}. 

About the value $T_{eff}$ in expression (\ref{eq_12}), we will need to know only the time interval for determining the stock’s return. If it is limited by the characteristic time of $G_S$ autocorrelation decrease only, we can assume with good accuracy that $T_{eff} (t + \Delta t) \simeq T_{eff} (t)$, and the most probable stock return $Y_n$ over the limited time interval $\Delta t$ will be determined by the expression

\begin{equation}
Y_n \simeq ln ({{\alpha}_{t + \Delta t} \over {\alpha}_t}) ,
	\label{eq_13}
\end{equation}

Let us now determine the ratio in expression (\ref{eq_13}), using the results of the previous section. 

The graphs showing the dependence of the value $\alpha$ on $m$ and $p_i$ (directly on ${p_i \over N}$) are presented in Figs. 1 and 2:

\begin{figure}
\includegraphics[width=1.0\linewidth]{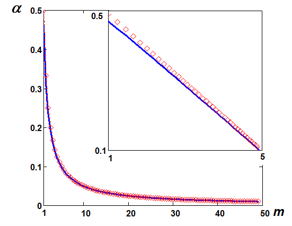}
\caption{\label{Fig.1} The dependence of $\alpha$ from the total number of portfolios $m$ containing the considered stock, in accordance with the second equation (\ref{eq_10}) – solid curve, uniform scale. Approximation (\ref{eq_14}) – open rhombs. The inset shows the same for small $m$, double logarithmic scale.}
\end{figure}

\begin{figure}
\includegraphics[width=1.0\linewidth]{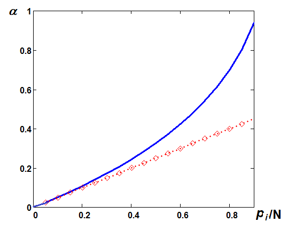}
\caption{\label{Fig.2} The dependence of $\alpha$ on the stock from the part of stocks in one portfolio ${p_i \over N}$. The solid curve represents the dependence according to the first equation (\ref{eq_10}), the open rhombs represent the dependence (\ref{eq_15}).}
\end{figure}

In reality, for the tremendous majority of stocks of various companies traded on the stock exchange, the relationships $m \gg 1$ and ${p_i \over N} \ll 1 $ hold (for other situations involving $m$ and ${p_i \over N}$, see ~\cite{romanovsky2026statistics}). Expanding the second equation (\ref{eq_10}) in powers of ${1 \over m}$, we have

\begin{equation}
{\alpha} \simeq - {1 \over{2m}}.
	\label{eq_14}
\end{equation}

In turn, the first equation (\ref{eq_10}) will give

\begin{equation}
{\alpha} \simeq - {p_i \over{2N}},
	\label{eq_15}
\end{equation}

i.e., in this approximation, we have $ {{m p_i} \over N} \simeq 1$ instead of the first equation (\ref{eq_10}), since $\alpha$ is very small for large $m$ and small  ${p_i \over{N}}$. Fig.1 shows that the dependence of the $\alpha$ value from the number of portfolios $m$ almost immediately converges to the approximation (\ref{eq_14}) already for relatively small $m$. The dependence ${p_i \over{N}}$ converges to the approximation (\ref{eq_15}) when ${p_i \over{N}} \le 1$, Fig.2. Thus, the figures indicate that approximations(\ref{eq_14})  and (\ref{eq_15})  practically work in all real cases of company stocks trading on the stock exchange. 

Expressions (\ref{eq_14}) and (\ref{eq_15}) (and the original equations (\ref{eq_10})) differ in meaning. (\ref{eq_14}) is essentially an algebraic equation. Expression (\ref{eq_15}) actually contains a random variable — the number of stocks in a certain portfolio, so (\ref{eq_15}) (and the first equation \ref{eq_10}) should be understood primarily as a relationship for some “average” $p_i$. For our purposes, relationship (\ref{eq_14}) is sufficient only. 

Taking the ratio of the values (\ref{eq_14}) at time $t$ and at time $t+\Delta t$ and substituting them into (\ref{eq_13}), we obtain the desired relationship for the return

\begin{equation}
Y_n \simeq - ln ({m_{t + \Delta t} \over m_t}) ,
	\label{eq_16}
\end{equation}

Here, $m_t$ is the number of portfolios containing the stocks in question at time t. Expression (\ref{eq_16}) solves the problem for the combinatorial determination of the most probable short term stock return over a (sufficiently short) time interval $\Delta t$.

Experimental confirmation of the developed theory is very difficult due to the practically complete impossibility of finding data on the current number of stockholders of a particular company \footnote{Formally, the number of stockholders is determined only for AGM, which may be attended by all stockholders.}. Since (\ref{eq_16}) includes  the ratio ${m_{t + \Delta t} \over m_t}$ only, some approximation can be obtained if there exist representative data for $m$ — for example, if the broker works with a large number of stock exchange trading participants $m'$, and we possess this information. Then the current ratio ${m'_{t+\Delta t} \over m'_t}$ can, within a certain tolerance, be considered equal to the true current ratio ${m_{t+\Delta t} \over m_t}$, and $ Y \simeq Y'_n \simeq - ln ({m'_{t + \Delta t} \over m'_t}) $.  

The developed theory can be experimentally verified by determining the correlation coefficient between $Y$ and $Y'_n$. In reality, information about the number of stockholders $m$ at a given time spreads with a certain speed, so we can expect that the most probable value of the return $Y$ will inevitably lag behind $Y'_n$. Therefore, it is necessary to calculate the correlation coefficients between $Y'_n$ and $Y$ at the time when $Y'_n$  is determined, as well as over a certain time interval after that moment. If the correlations between the values of $Y$ and $Y'_n$ will be sufficiently high (the criterion can be taken as a correlation coefficient greater than 0.5), then the developed method can apparently be used to determine (predict) stock prices in order to make recommendations regarding to buy or sale the stock. 

Fig.3 shows the correlation coefficients that were calculated in the period March–April 2026. Quadrats show the correlation coefficient of SBERBANK ordinary stocks one-day experimental return with the one-day $Y'_n$ (i.e. $\Delta t = 24$ hours for both, and $t$ is yesterday). Rings show correlation coefficient of LUKOIL stocks two-day experimental return with the two-day $Y'_n$ (i.e. $\Delta t =48$ hours, and $t$ is the day before yesterday) . One can see that there are no correlations on the same day (the correlation coefficient is less than zero). In front of that, they increase sharply within two hours the next morning, i.e. the delay of $Y$ with respect to $Y'_n$ is several hours. Thus if $Y'_n$, as in our case, was determined in the evening (see Fig.3), then the correlation maximum is realized in the morning of the next day after the opening within an hour.

\begin{figure}
\includegraphics[width=1.0\linewidth]{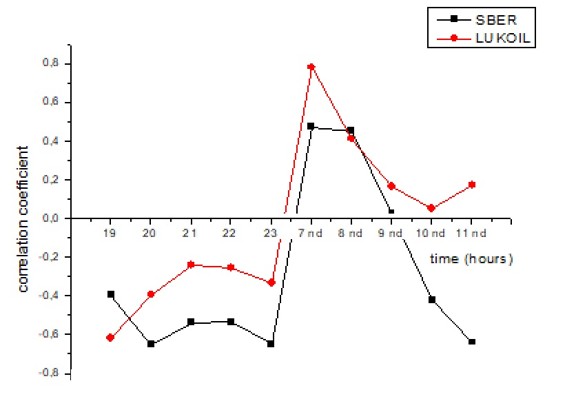}
\caption{\label{Fig.3} Correlation coefficients of the one day return of SBERBANK ordinary stocks with the one day value $- ln ({m'_{t + \Delta t} \over m'_t})$ (dark squares connected by solid lines), calculated for the corresponding hour of the day.
 Correlation coefficients of the two day return of LUKOIL stocks with the two day value $- ln ({m'_{t + \Delta t} \over m'_t})$ (dark circles connected by solid lines), also calculated for the corresponding hour of the day. The value $m'_t$ was determined at 18:00 each day, the notation "nd" indicates the new (next) day.}
\end{figure} 

Thus, for example, if the quantity $m'_{t+ \Delta t}$ falls with respect to $m'_t$, a stock market player has time in the evening to buy stocks and, early the next morning, sell them and make a profit.

\section{Final remarks. }

We have examined a combinatorial approach to the problem of determining the return of stocks (over a short time interval) and have identified its most probable value. It would seem that the result obtained for a single company is of limited scope, since the calculations did not directly take into account the influence of the entire stock market. However, if we take into account the efficient market hypothesis (see, for example, \cite{mantegna2000stanley}), all information about the market (not about the considered company itself) has already been processed in the values of $p_i$ and $m$. The speed of such processing has significantly increased over the past 10 years due to the widespread use of corresponding robot brokers for trading, which, unlike human brokers, process arbitrage opportunities almost instantly. Thus, even approximate expressions (\ref{eq_14}), (\ref{eq_15}) can be considered instantaneous most probable values, and for limited time intervals, profitability can be determined using formula (\ref{eq_16}). 

The possibilities for forecasting stock prices taking into account (\ref{eq_16}) are clear — if the current return differs from (\ref{eq_16}) in one direction or another, the recommendations for buying or selling stocks are obvious: if the current yield determined over a time interval is less than the yield calculated using formula (\ref{eq_16}), it is advisable to buy the stocks in question. If the opposite is true, then sell. The inevitable delay in the maximum of the correlation coefficient compared to the time when $Y_n$ (or $Y'_n$) is determined gives time to make such decisions.

\section{Acknowledgments.} 

The author thanks I.Zhukov and A.Shishov for their great support in the work and for the fruitful discussion of the results.


\begin{thebibliography}{38}
\bibitem{isihara1971statistical} A.Isihara, Statistical Physics  (Academic Press, New York - London, 1971).

\bibitem{romanovsky2021statistical} M.Yu.Romanovsky, Statistical properties of many-particle systems with hierarchial structute, Journal of Experimental and Theoretical Physics (JETP) {\bf 132}, 294 (2021). 

\bibitem{romanovsky2022chaos} M.Yu.Romanovsky, Most probable distributions and distributions of extremes for particle systems with hierarchical structures, Chaos, Solitons and Fractals. {\bf 159}, 112170 (2022)

\bibitem{romanovsky2026UFN} M.Yu.Romanovsky, The exclusion principle and the evolution of matter in statistics of particles with hierarchical structure, Phys. Usp. {\bf 69}, 714 (2026)

\bibitem{romanovsky2011vidov} M.Yu.Romanovsky, P.V.Vidov, Analytical representation of stock and stock-indexes returns: Non-Gaussian random walks with various jump laws, Physica A {\bf 390}, 3794 (2011) 

\bibitem{mantegna2000stanley}  R.N.Mantegna, H.E.Stanley, An Introduction to Econophysics. Correlation and Complexity in Finance, (Cambridge University Press, NY, 1997).

\bibitem{romanovsky2020romanovsky} M.Yu.Romanovsky, Yu.M.Romanovsky, Mathematical Principles of Econophysics [Matematicheskie nachala ekonofiziki]  (Institute of Computer Science, Moscow–Izhevsk 2020) [in Russian].

\bibitem{stanley1999} Y.Liu, P.Gopikrishnan, P.Cizeau, M.Meyer, C-K.Peng, and H.E.Stanley, Statistical properties of the volatility of price fluctuations. Phys.Rev.E., {\bf 60}, 1390 (1999). 

\bibitem{romanovsky2026statistics} M.Yu.Romanovsky, Statistical Physics. Additional Parts [Statisticheskaya fizika. Dopolnitel'nie glavy] (Institute of Computer Science, Moscow–Izhevsk 2026) [in Russian].

\end{thebibliography}

\end{document}